\documentclass{aa}
\usepackage{psfig}
\usepackage{latexsym}

\newcommand{\MC}{\multicolumn}
\newcounter{qub}
\begin{document}

\title{On the metallicities of UM~133, UM~283 and UM~382}

\author{%
Kniazev A.Y.\inst{1}\fnmsep\inst{2}
\and Pustilnik S.A.\inst{1}\fnmsep\inst{2}
\and Ugryumov A.V.\inst{1}\fnmsep\inst{2}
\and Pramsky A.G.\inst{1}
}

\offprints{Kniazev A.Y., \email{akn@sao.ru}}

\institute{
Special Astrophysical Observatory, Nizhnij Arkhyz, Karachai-Circassia,
369167, Russia \\
\and Isaac Newton Institute of Chile, SAO Branch
}

\date{Received January 30, 2000; accepted February 10, 2001}

\markboth{Kniazev et al.: On the metallicities of UM~133, UM~283 and UM~382}
{On the metallicities of UM~133, UM~283 and UM~382}

\abstract{ {
The study of group properties of the extremely metal-deficient gas-rich local
dwarfs is very promising for the understanding the galaxy formation process
at high redshifts. About 20 such objects have been picked up from the
literature in the recent review by Kunth \& \"Ostlin~(\cite{Kunth00}).
However part of these galaxies got low metallicity as a result of earlier
observations, and can have rather large uncertainties in their cited element
abundances.
Before to perform the detailed studies of such galaxies as of some extreme
group, it is useful to revise their metallicities.}
We present the results of the SAO 6\,m telescope spectrophotometry of two
Blue Compact Galaxies (BCG) reported from earlier studies as  very metal-poor
objects. Well measured [O~{\sc iii}] line $\lambda$4363 \AA\
allows to deduce the temperature in H{\sc ii} regions and get reliable
abundances of chemical elements.
For UM~133 we derive 12+log(O/H) = 7.63$\pm$0.02, coincident with the
published value.
UM~382, according to our data, is significantly more metal-rich: its
12+log(O/H) = 7.82$\pm$0.03 in comparison to the published value 7.45.
The third galaxy, UM~283 seems have got its very low 12+log(O/H) = 7.59 due
to a misprint. We used its published emission line intensities  and
derived instead the value of 7.95. Thus the latter two galaxies should
{\it NOT} be considered as the extremely metal-poor BCGs.
}

\maketitle

% --------------------------------------
\section{Introduction}
% --------------------------------------

Since the discovery of low metallicity gas in H {\sc ii} regions of
Blue Compact Galaxies (BCGs) by Searle \& Sargent (\cite{Searle72}),
a debate lasting for
decades on possible existence of local truly young galaxies is still
continuing. While the great majority of BCGs have metallicities Z in the range
of (1/10 -- 1/3) Z$_\odot$, or respectively 12+log(O/H) in the range
7.92 -- 8.42 \footnote{12+log(O/H)$_{\odot}$ = 8.92
(Anders \& Grevesse \cite{Anders89}).}, very few galaxies have O-abundance
as low as 12+log(O/H) = 7.1--7.6, consistent with the expected oxygen
enrichment produced in one star formation (SF) episode. Izotov \& Thuan
(\cite{IT99}) from the analysis of variations of C and N abundances have
shown, that namely the BCGs with 12+log(O/H) $\leq$ 7.6 are the best
candidates to the galaxies with the first SF episode. To clear up the true
evolution status of these extremely metal-poor BCGs is a serious challenge
for observational astrophysics. However, independently on whether their
current SF burst is the first or second one, properties of these local well
resolved galaxies best approximate those of primeval low-mass galaxies at
high redshifts.
Therefore systematic studies of the group properties of this small
sample are very important from cosmological point of view.

Kunth \& \"Ostlin (\cite{Kunth00}) (K\"O), in their review of the problems
related to the most metal-poor galaxies, summarized published up-to-now data
and presented the compilative list of these galaxies. In course of the
studies of new large samples of BCGs in a broader context, we attempt to find
and study as well new very metal-poor galaxies (e.g., Kniazev et
al.~\cite{Kniazev98}, \cite{Kniazev2000a}, \cite{Kniazev2000b}, Pustilnik et
al.~\cite{VLA}). In parallel we check old
measurements for several BCGs, claimed to have very low metallicities.
In particular, we reobserved two BCGs presented in the K\"O list --
UM~133 and UM~382. Reticon spectrophotometry for these two BCGs is
originally from Terlevich et al. (\cite{TMMMC91}) and Masegosa et al.
(\cite{MMMC-A94}).

In this paper we present a new spectrophotometry with the SAO RAS 6\,m
telescope for UM~133 and UM~382.
For UM~283 with its published
emission line intensities we revised its claimed very low O abundance.
Hubble constant $H_\mathrm{0}$ = 75~km s$^{-1}$Mpc$^{-1}$
was used through the paper.

%&&&&&&&&&&&&&&&&&&&&&&&&&&&&&&&&&&&&&&&&&&&&&&&&&&&&&
% Table 1. General parameters
%&&&&&&&&&&&&&&&&&&&&&&&&&&&&&&&&&&&&&&&&&&&&&&&&&&&&&
\begin{table*}
\begin{center}
\caption{\label{Tab1} General Parameters of Studied Galaxies}
\begin{tabular}{llrllrll} \\ \hline \hline
\MC{1}{c}{ Galaxy }            &
\MC{1}{c}{ $\alpha_\mathrm{2000.0}$ } &
\MC{1}{c}{ $\delta_\mathrm{2000.0}$ } &
\MC{1}{c}{ B$_\mathrm{tot}$ }         &
\MC{1}{c}{ $(B-V)_\mathrm{tot}$}      &
\MC{1}{c}{ A$_\mathrm{B}^\mathrm{N}$ }         &
\MC{1}{c}{ $V_\mathrm{hel}$ }         &
\MC{1}{c}{ $M_\mathrm{B}^\mathrm{E,1}$ }     \\

\MC{1}{c}{ Name }              &
\MC{1}{c}{ $h$~$m$~$s$ }       &
\MC{1}{c}{ $\circ$~$\prime$~$\prime\prime$ } &
\MC{1}{c}{ $mag$ }             &
\MC{1}{c}{ $mag$ }             &
\MC{1}{c}{       }             &
\MC{1}{c}{ $km/s$ }            &
\MC{1}{c}{ $mag$ }             \\

\MC{1}{c}{ (1) } &
\MC{1}{c}{ (2) } &
\MC{1}{c}{ (3) } &
\MC{1}{c}{ (4) } &
\MC{1}{c}{ (5) } &
\MC{1}{c}{ (6) } &
\MC{1}{c}{ (7) } &
\MC{1}{c}{ (8) } \\
\hline
\\[-0.3cm]
UM~133  & 01 44 41.3 & +04 53 26 & 15\fm71$\pm$0\fm05$^1$ & 0.30$\pm$0.06$^{1,4}$ &0.16& 1623$\pm$4$^2$  &--16\fm04  \\
UM~283  & 00 51 49.5 & +00 33 53 & 16\fm97$\pm$0\fm02$^5$ &                       &0.10& 4197$^5$        &--16\fm67  \\
UM~382  & 01 58 09.4 &--00 06 38 & 18\fm56$\pm$0\fm06$^3$ & 0.40$\pm$0.08$^3$     &0.12& 3526$\pm$30$^0$ &--14\fm91  \\
\hline \\[-0.2cm]
\MC{8}{l}{B$_\mathrm{tot}$ -- total blue magnitude; A$_\mathrm{B}$ -- Galactic extinction; M$_\mathrm{B}$ -- absolute blue magnitude;} \\
\MC{8}{l}{$^\mathrm{N}$\,\ Data from NED; $^\mathrm{E}$\,\ With Galactic extinction correction;} \\
\multicolumn{8}{l}{{\bf References:} $^0$ This paper $^1$ Kniazev et al.~(\cite{UM133}); $^{2}$ Thuan et al.~(\cite{Thuanetal99}); $^3$ Salzer et al.~(\cite{SAB89});}\\
\multicolumn{8}{l}{$^4$ Telles \& Terlevich~(\cite{TT97}); $^5$ P\'{e}rez-Gonz\'{a}les et al. (\cite{PG2000}).}
\end{tabular}
\end{center}
\end{table*}

%&&&&&&&&&&&&&&&&&&&&&&&&&&&&&&&&&&&&&&&&&&&&&&&&&&&&&
% Table 2. Obs. log
%&&&&&&&&&&&&&&&&&&&&&&&&&&&&&&&&&&&&&&&&&&&&&&&&&&&&&
\begin{table*}
\begin{center}
\caption{\label{Tab2} Journal of Observations}
\begin{tabular}{lllrcccccc} \\ \hline \hline
\MC{1}{c}{ Galaxy }     &
\MC{1}{c}{ Date }       &
\MC{1}{c}{ Instrument } &
\MC{1}{c}{ Exposure }   &
\MC{1}{c}{ Wavelength } &
\MC{1}{c}{ Dispersion } &
\MC{1}{c}{ Slit }       &
\MC{1}{c}{ Seeing }     &
\MC{1}{c}{ PA }         &
\MC{1}{c}{ Airmass } \\

\MC{1}{c}{ Name }        &  &  &
\MC{1}{c}{ time [s] }    &
\MC{1}{c}{ Range [\AA] } &
\MC{1}{c}{ [\AA/pixel] } &
\MC{1}{c}{ [arcsec] }    &
\MC{1}{c}{ [arcsec] }    & & \\

\MC{1}{c}{ (1) } &
\MC{1}{c}{ (2) } &
\MC{1}{c}{ (3) } &
\MC{1}{c}{ (4) } &
\MC{1}{c}{ (5) } &
\MC{1}{c}{ (6) } &
\MC{1}{c}{ (7) } &
\MC{1}{c}{ (8) } &
\MC{1}{c}{ (9) } &
\MC{1}{c}{ (10) } \\
\hline
\\[-0.3cm]
UM~133        & 04.09.1999 & 6\,m, LSS & 1200~~~~ & $ 3600-8000$  & 4.6 & 1.3 & 1.5 & 22 & 1.31 \\
UM~133        & 03.10.2000 & 6\,m, LSS & 3600~~~~ & $ 3700-6100$  & 2.4 & 2.0 & 1.2 & 22 & 1.30 \\
UM~133        & 05.10.2000 & 6\,m, LSS & 1800~~~~ & $ 5800-8200$  & 2.4 & 2.0 & 1.2 & 22 & 1.29 \\
UM~382        & 04.09.1999 & 6\,m, LSS & 1200~~~~ & $ 3600-8000$  & 4.6 & 1.3 & 1.5 & 46 & 1.52 \\
UM~382        & 06.10.1999 & 6\,m, LSS & 1200~~~~ & $ 3700-6100$  & 2.4 & 2.0 & 1.0 & 46 & 1.39 \\
UM~382        & 03.10.2000 & 6\,m, LSS & 1800~~~~ & $ 3700-6100$  & 2.4 & 2.0 & 1.2 &  0 & 1.39 \\
UM~382        & 05.10.2000 & 6\,m, LSS & 1800~~~~ & $ 5800-8200$  & 2.4 & 2.0 & 1.2 &  0 & 1.39 \\
\hline \\[-0.2cm]
\end{tabular}
\end{center}
\end{table*}

%*************************************************************
% Fig.1  Direct image of UM~133 and 1D spectrum of UM~133
%*************************************************************
\begin{figure*}
{\centering
\hspace*{0.0cm}
\psfig{figure=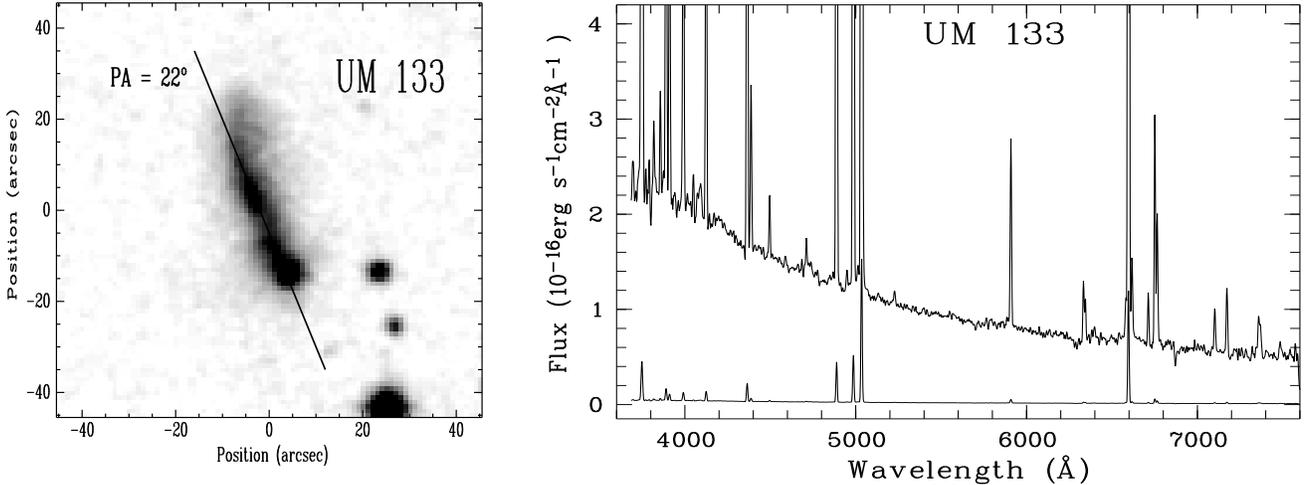,angle=0,width=18cm,clip=,bbllx=35pt,bblly=325pt,bburx=575pt,bbury=530pt}
}
\caption{
{\it Left panel:}
Digitized Sky Survey (DSS-2) blue image of UM~133 with the position of long
slit superimposed.
The bright star-forming complex is visible at the SW end of the elongated
host. On the accepted distance 22.4 Mpc 1\arcsec\ corresponds to 108 pc.
{\it Right panel:}
1-D spectrum of the bright SW star-forming region in
UM~133  extracted from 2-D spectrum observed with dispersion 2.4~\AA/pixel.
The lower one is scaled by the factor of 1/50 to show
the relative intensities of strong lines.
}
\label{UM133_direct_fig}
\end{figure*}

%*************************************************************
% Fig.2  Direct image of UM~382 and 1D spectrum of UM~382
%*************************************************************
\begin{figure*}
{\centering
\psfig{figure=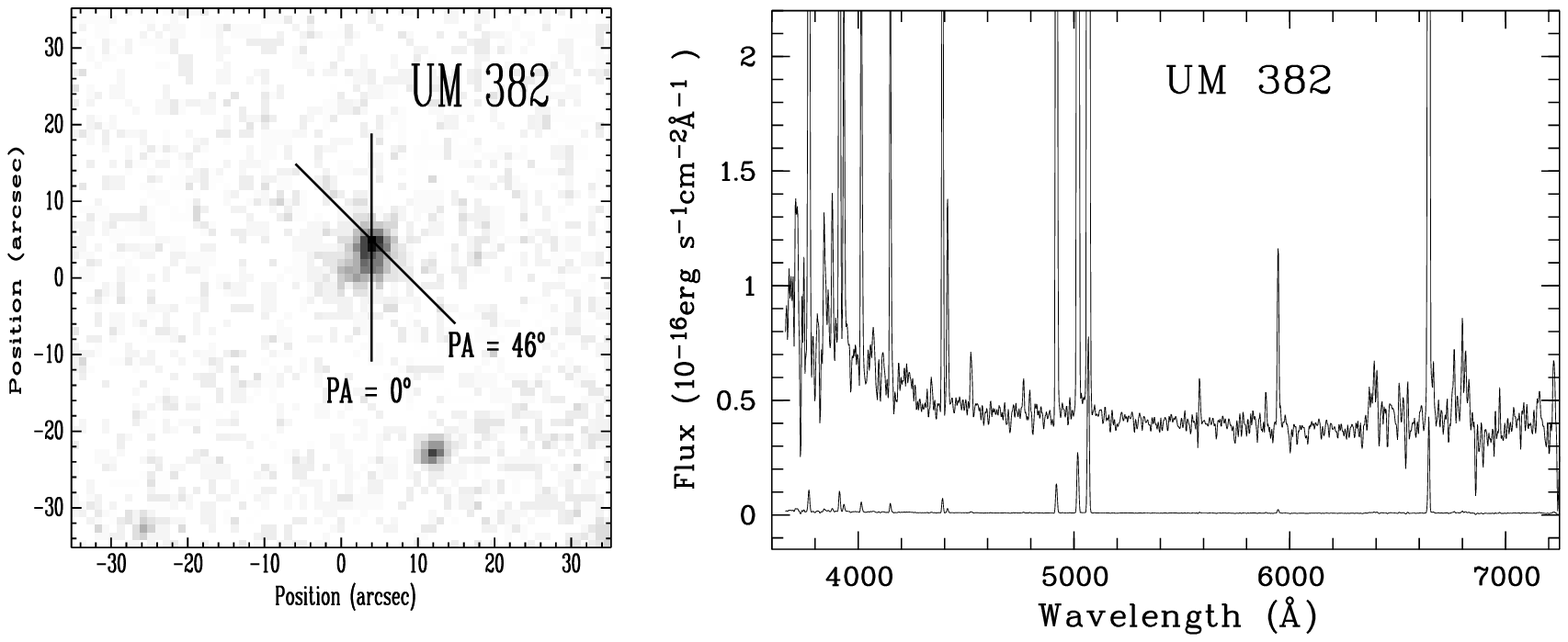,angle=0,width=18cm,clip=,bbllx=35pt,bblly=120pt,bburx=555pt,bbury=325pt}
}
\caption{{\it Left panel:}
DSS-2 red image of UM~382 with the long slit position superimposed.
{\it Right panel:} 1-D spectrum of UM~382 with dispersion 2.4~\AA/pixel;
the lower one is scaled by factor of 1/10 to show the relative intensities
of strong lines.
}
\label{UM382spec_fig}
\end{figure*}

% --------------------------------------
\section{Observations and data reduction}
% --------------------------------------

Some parameters of the studied galaxies either known from the literature or
derived in this paper are presented in
Table~\ref{Tab1}.
The spectroscopic data were obtained  with the
6\,m telescope of the Special Astrophysical Observatory of Russian
Academy of Science (SAO RAS) during the runs in September 1999 and October
2000 (see Table~\ref{Tab2} for details).
The Long-Slit spectrograph (LSS) (Afanasiev et al. \cite{Afanasiev95})
at the telescope prime focus was equipped with the
Photometrics CCD-detector  1024$\times$1024 pixels with
$24\times24\mu$m pixel size.
The long-slit (180\arcsec)  spectra were obtained
with the gratings of 325 and 650 grooves/mm with corresponding
dispersions of 4.6 and 2.4~\AA/pixel
and a spectral resolution 12-15~\AA\ (FWHM) and 7-9~\AA\ respectivelty.
The wavelength ranges of the obtained spectra for different hardware
configurations are given in Table~\ref{Tab2}.
The slit widths of 1.3\arcsec\ and 2\arcsec\ were used.
The scale along the slit was 0.39~\arcsec/pixel in all cases.
For UM~133 the slit, centered on the brightest SW knot (see in
Fig.~\ref{UM133_direct_fig}),
was oriented along the main body of the galaxy.

Reference spectra of an Ar--Ne--He lamp were recorded before or after
each observation to provide  wavelength calibration.
Spectrophotometric standard stars from Bohlin (\cite{Bohlin96})
were observed for flux calibration.
Observations have been
conducted mainly under the software package {\tt NICE} in MIDAS, described by
Kniazev \& Shergin (\cite{Kniazev95}).

{To be confident that we observed the same parts of the galaxies with the
different set-ups in different nights we employed the differential method of
pointing the telescope to the specific position in the target galaxy. It
consisted of the following steps:
1). the positions of both the specific bright knot in the target galaxy and
that for off-set bright enough star at the distance $<$(1--2)$^{\prime}$ were
carefully measured from DSS-2;
2). the position angle of the slit for observations of the target was set;
3). the telescope was pointed on the coordinates of the off-set star;
4). the center of the slit was shifted to the center of this off-set star
    and the respective correction was saved by the telescope pointing program;
5). finally the telescope was pointed on the coordinates of the specific
    knot in the target galaxy using this correction. The latter was
    defined for each observation independently, since it depends on concrete
    azimuth and zenith distance of the target.

Procedures of primary data reduction included cosmic-ray removal
in MIDAS\footnote{MIDAS is an acronym for the European Southern Observatory
package -- Munich Image Data Analysis System.}
and bias subtraction and flat-field correction in
IRAF\footnote{IRAF: the Image Reduction and Analysis Facility is
distributed by the National Optical Astronomy Observatories, which is
operated by the Association of Universities for Research in Astronomy,
In. (AURA) under cooperative agreement with the National Science
Foundation (NSF).} software packages.
For the following reduction of the long-slit spectra we used IRAF.
After the wavelength mapping and night sky subtraction,
each 2D frame was corrected for atmospheric extinction and was flux
calibrated.
To derive the sensitivity curves, we used the
spectral energy distributions of the standard stars.
Average sensitivity curves were produced for each observing night.
2D blue and red parts of spectra were extracted using {\tt APALL} IRAF task
and 2D combined spectrum for each object was created by {\tt SCOMBINE} task.

Finally, the 1D spectrum was extracted from a
region along the slit, where
\mbox{$I$($\lambda$4363~\AA)} $>$ 2$\sigma$ ($\sigma$ is the dispersion of
a noise statistics around this line).
The size of integrated region was 3.5\arcsec\ for UM~133 and
3.1\arcsec\ for UM~382.
The 1D extracted spectra of UM~133 and UM~382 are shown in
Fig.~\ref{UM133_direct_fig} and Fig.~\ref{UM382spec_fig}.

%&&&&&&&&&&&&&&&&&&&&&&&&&&&&&&&&&&&&&&&&&&&&&&&&&&&&&
% Table 3. Intensities of lines
%&&&&&&&&&&&&&&&&&&&&&&&&&&&&&&&&&&&&&&&&&&&&&&&&&&&&&
\begin{table*}[hbtp]
\centering{
\caption{Line intensities of the studied galaxies}
\label{t:Intens}
\begin{tabular}{lcccccc} \hline
\rule{0pt}{10pt}
					  & \MC{2}{c}{UM~133}                     &  \MC{2}{c}{UM~382} &  \MC{2}{c}{UM~283$^\mathrm{\bf b}$}                     \\ \hline
$\lambda_\mathrm{0}$(\AA) Ion         & F($\lambda$)/F(H$\beta$)&I($\lambda$)/I(H$\beta$)& F($\lambda$)/F(H$\beta$)&I($\lambda$)/I(H$\beta$) & F($\lambda$)/F(H$\beta$)&I($\lambda$)/I(H$\beta$) \\ \hline
3727\ [O\ {\sc ii}]\           & 1.1712$\pm$0.0148 & 1.1889$\pm$0.0159 &  0.689$\pm$0.013 & 0.784$\pm$0.038 & 5.24 & 5.74      \\
3835\ H9\                      & 0.0492$\pm$0.0040 & 0.0614$\pm$0.0065 &  0.060$\pm$0.002 & 0.072$\pm$0.007 & ---  & ---       \\
3868\ [Ne\ {\sc iii}]\         & 0.3229$\pm$0.0072 & 0.3268$\pm$0.0074 &  0.678$\pm$0.022 & 0.758$\pm$0.038 & ---  & ---       \\
3889\ He\ {\sc i}\ +\ H8\      & 0.1799$\pm$0.0066 & 0.1930$\pm$0.0082 &  0.267$\pm$0.017 & 0.300$\pm$0.022 & ---  & ---       \\
3967\ [Ne\ {\sc iii}]\ +\ H7\  & 0.2319$\pm$0.0060 & 0.2450$\pm$0.0074 &  0.253$\pm$0.010 & 0.283$\pm$0.016 & ---  & ---       \\
4101\ H$\delta$\               & 0.2523$\pm$0.0079 & 0.2642$\pm$0.0090 &  0.255$\pm$0.010 & 0.280$\pm$0.014 & ---  & ---       \\
4340\ H$\gamma$\               & 0.4627$\pm$0.0065 & 0.4725$\pm$0.0074 &  0.442$\pm$0.010 & 0.468$\pm$0.015 & 0.39 & 0.41      \\
4363\ [O\ {\sc iii}]\          & 0.0869$\pm$0.0046 & 0.0872$\pm$0.0046 &  0.138$\pm$0.006 & 0.146$\pm$0.007 & ---  & ---       \\
4686\ He\ {\sc ii}\            & 0.0118$\pm$0.0032 & 0.0118$\pm$0.0033 &  ---             & ---             & ---  & ---       \\
4861\ H$\beta$\                & 1.0000$\pm$0.0127 & 1.0000$\pm$0.0130 &  1.000$\pm$0.025 & 1.000$\pm$0.025 & 1.00 & 1.00      \\
4959\ [O\ {\sc iii}]\          & 1.2134$\pm$0.0123 & 1.2039$\pm$0.0123 &  2.223$\pm$0.050 & 2.197$\pm$0.050 & 1.15 & 1.14      \\
5007\ [O\ {\sc iii}]\          & 3.6418$\pm$0.0352 & 3.6105$\pm$0.0351 &  6.450$\pm$0.130 & 6.346$\pm$0.130 & 3.24 & 3.21      \\
6563\ H$\alpha$\               & 2.8514$\pm$0.0284 & 2.7725$\pm$0.0302 &  3.214$\pm$0.344 & 2.787$\pm$0.325 & 3.15 & 2.86      \\
6717\ [S\ {\sc ii}]\           & 0.1148$\pm$0.0053 & 0.1113$\pm$0.0052 &  0.091$\pm$0.016 & 0.078$\pm$0.014 & 0.42 & 0.38      \\
6731\ [S\ {\sc ii}]\           & 0.0674$\pm$0.0048 & 0.0654$\pm$0.0047 &  0.066$\pm$0.016 & 0.057$\pm$0.014 & 0.39 & 0.35      \\
  & & \\
C(H$\beta$)\ dex          & \MC {2}{c}{0.03$\pm$0.01} & \MC {2}{c}{0.18$\pm$0.14} & \MC{2}{c}{---}\\
EW(abs)\ \AA\             & \MC {2}{c}{1.00$\pm$0.35} & \MC {2}{c}{0.25$\pm$0.35} & \MC{2}{c}{---}          \\
F(H$\beta$)$^\mathrm{\bf a}$\           & \MC {2}{c}{191$\pm$2} & \MC {2}{c}{60$\pm$2}         & \MC{2}{c}{136}         \\
EW(H$\beta$)\ \AA\        & \MC {2}{c}{ 159$\pm$1}     & \MC {2}{c}{ 135$\pm$2}   & \MC{2}{c}{122}         \\
\hline
\MC{7}{l}{~~} \\
\MC{7}{l}{$^\mathrm{\bf a}$ in units of 10$^{-16}$ ergs\ s$^{-1}$cm$^{-2}$; ~~~$^\mathrm{\bf b}$ all data are taken from Gallego et al.~(\cite{Gallego96}).}\\
\end{tabular}
}
\end{table*}

All spectra of UM~133 and UM~382 obtained during the different observational
runs (see Table~\ref{Tab2}) were used for calculation of chemical element abundances.
These values of abundances are consistent within the observational
uncertainties for different runs.
As final results we used the best quality spectra obtained during
October 2000 run.}

The redshifts and line fluxes were measured with MIDAS software as described
in Kniazev et al. (\cite{Kniazev2000b}) and Hopp et al. (\cite{Hopp2000}).
The errors of the line intensities have been propagated in the calculations
of the element abundances.
For the simultaneous derivation  of $C$(H$\beta$) and $EW$(abs)  and
correction for extinction we used the procedure described in detail
by Izotov et al. (\cite{Izotov97b}).
The high S/N ratio 6\,m spectra permit to derive the element abundances
with a higher precision than in the previous studies.
The abundances of the ionized species and the total abundances 
of O, Ne and Ar also have been obtained following the
procedure detailed in Izotov et al. (\cite{Izotov94}),
Thuan et al. (\cite{Thuan95}), Izotov et al. (\cite{Izotov97b}.
The abundance of O$^{3+}$ ion is derived following to
Izotov \& Thuan (\cite{IT99})  with use of
the intensity of \ion{He}{ii} $\lambda$4686 line.

The observed emission line intensities $F(\lambda)$, and those corrected
for the interstellar extinction and underlying stellar absorption $I(\lambda)$
are presented in Table~\ref{t:Intens}.
The observed H$\beta$ equivalent width $EW$(H$\beta$),
absorption Balmer hydrogen lines equivalent widths $EW$(abs),
H$\beta$ flux and the extinction coefficient $C$(H$\beta$) are also shown
in Table~\ref{t:Intens}.

%&&&&&&&&&&&&&&&&&&&&&&&&&&&&&&&&&&&&&&&&&&&&&&&&&&&&&
% Table 4. Abundances
%&&&&&&&&&&&&&&&&&&&&&&&&&&&&&&&&&&&&&&&&&&&&&&&&&&&&&
\begin{table}[hbtp]
\centering{
\caption{Abundances in studied galaxies}
\label{t:Chem}
\begin{tabular}{lccc} \hline
\rule{0pt}{10pt}
Value                                & UM 133             &  UM 382 &  UM 283\\ \hline
$T_\mathrm{e}$(OIII)(K)              & 16,669$\pm$451   &  16,187$\pm$421  &  ~16,100$^\mathrm{\bf a}$      \\
$T_\mathrm{e}$(OII)(K)               & 14,502$\pm$373   &  14,297$\pm$354  &  14,070                \\
$T_\mathrm{e}$(SIII)(K)              & 15,535$\pm$374   &  15,135$\pm$349  &     ---                \\
$N_\mathrm{e}$(SII)(cm$^{-3}$)       & $<$10            &   45$\pm$376     &  100$^\mathrm{\bf a}$         \\
& \\
O$^{+}$/H$^{+}$($\times$10$^5$)      & 1.140$\pm$0.080  &  0.789$\pm$0.065 &  2.99      \\
O$^{++}$/H$^{+}$($\times$10$^5$)     & 3.055$\pm$0.204  &  5.800$\pm$0.387 &  6.09      \\
O$^{+++}$/H$^{+}$($\times$10$^5$)    & 0.063$\pm$0.020  &  ---             &  ---       \\
O/H($\times$10$^5$)                  & 4.258$\pm$0.220  &  6.589$\pm$0.392 &  9.08      \\
12+log(O/H)                          & ~7.63$\pm$0.02   &  ~7.82$\pm$0.03  &  7.95      \\
& \\
Ne$^{++}$/H$^{+}$($\times$10$^5$)    & 0.594$\pm$0.043  &  1.490$\pm$0.128 &  ---      \\
ICF(Ne)$^\mathrm{\bf b}$             & 1.394            &  1.136           &  ---      \\
log(Ne/O)                            & --0.71$\pm$0.04  &  --0.59$\pm$0.05 &  ---      \\
\hline
\MC{4}{l}{~~} \\
\MC{4}{l}{$^\mathrm{\bf a}$ data from Gallego et al.~(\cite{Gallego97}).}\\
\MC{4}{p{8cm}}{$^\mathrm{\bf b}$ ICF is the ionization correction factor for unseen stages of
ionization. The expressions for ICFs are adopted from Izotov et al. (\cite{Izotov94}).
}
\end{tabular}
 }
\end{table}

% --------------------------------------
\section{Results}
% --------------------------------------

The results of the chemical abundance determination for the studied galaxies
are presented in Table~\ref{t:Chem}.
The comparison of the respective data for Ne abundance for UM~133 and UM~382
with the abundance ratios from Izotov \& Thuan (\cite{IT99}) shows that
they well agree with the derived average values
for the sample of low metallicity BCGs.

% --------------------------------------
\subsection{UM~133}
% --------------------------------------

The galaxy is elongated and resembles on morphology a comet-like object.
In fact, according to NED, UM~133 is a bright H~{\sc ii} region at the
SW edge of Sc galaxy CGCG 412--024.
It looks like an edge-on disk, bent on NE edge.
The full range of the velocity curve from our 2D spectrum is about
100~km~s$^{-1}$.
This is slightly lower than W$_{0.2}$=122$\pm$11~km~s$^{-1}$ -- the full
width for the 21 cm line of the integrated H{\sc i} emission at the level
0.2 of peak (Thuan et al. \cite{Thuanetal99}).
We derived from our data the systemic radial velocity of host galaxy
of 1620$\pm$15~km~s$^{-1}$, which is well consistent with that
derived from the integral H{\sc i} profile (Thuan et al. \cite{Thuanetal99})
and with optical data from Huchra et al.~(\cite{Huchra1999}).
The radial velocity of UM~133 itself is on our data 1590 km~s$^{-1}$.
The continuous H$\alpha$-emission and a smooth velocity distribution
along the galaxy body supports its interpretation as a single
galaxy (CGCG~412--024) with a bright H{\sc ii} region (UM~133) at the SW edge.

As it is well seen on 1D spectrum
of UM~133, there is blue bump near $\lambda$4700\AA, characteristic of WR
stars. Its detailed quantitative analysis along with other observational data
is the subject of a forthcoming paper by Kniazev et al.~(\cite{UM133}).

% --------------------------------------
\subsection{UM~382}
% --------------------------------------

The galaxy has a star-forming region on the Northern edge. We got the
spectra of this region during 2 observational runs. The derived  oxygen
abundance of UM~382 (12+log(O/H) = 7.82$\pm$0.03) is significantly higher
than those presented by Masegosa et al.~(\cite{MMMC-A94})
(7.45$\pm$0.04) and derived on the same data by Telles~(\cite{Telles95})
(7.52$\pm$0.07). Since the latter results were obtained from
a Reticon spectrum with a lower S/N ratio and our data from two
observational runs are consistent each to other,
we consider the new data as a more reliable. Therefore
UM~382 is not an extremely metal-poor H~{\sc ii} galaxy.
The main reason for
the difference with earlier results is in the relative fluxes of [O{\sc iii}]
$\lambda$4363 line. Masegosa et al.~(\cite{MMMC-A94}) got for this parameter
the value of 0.25, while in our spectra it is only 0.138$\pm$0.006.
The another more reliable value from our data is the extinction coefficient
$C$(H$\beta$) = 0.18 in opposite to quite unusual  for this sort
of galaxies value of 0.71 from Terlevich et al.~(\cite{TMMMC91}).
The heliocentric velocity for this galaxy on our data is V$_\mathrm{hel}$ =
3526$\pm$30 km~s$^{-1}$ in comparison to the value of 3598 km~s$^{-1}$ from
Terlevich et al. (\cite{TMMMC91}).

% --------------------------------------
\subsection{UM~283}
% --------------------------------------

This galaxy have appeared with the oxygen abundance of 12+log(O/H)=7.59
in Gallego et al. (\cite{Gallego97}) as UCM~0049+0017. We have paid attention
that the relative
line intensities for this object reproduced in our Table~1 from Table~4 of
Gallego et al. (\cite{Gallego96}), according to the standard method by Pagel
et al. (\cite{Pagel92}) lead, with
T$_\mathrm{e}$=16,100 (Gallego et al.~\cite{Gallego97})
to the value 7.95, but not 7.59.
Therefore, we suggest that UM~283 appeared as a very metal-poor object just
as a result of a misprint.

% --------------------------------------
\section{Conclusions}
% --------------------------------------

From the data and discussion above we draw the following conclusions:
\begin{enumerate}
\item
UM~133 is confirmed as an object with very low metallicity
(Z $\approx$ 1/20 Z$_{\odot}$). From the radial velocity distribution
along the slit we conclude that this is a bright H{\sc ii} region on
the SW edge of dwarf comet-like galaxy CGCG~412--024. The radial velocities
of this H{\sc ii} region and the whole galaxy differ by no more than
40~km~s$^{-1}$.
\item
 UM~382 (Z $\sim$ 1/13 Z$_{\odot}$) and UM~283 (Z$\sim$ 1/9 Z$_{\odot}$)
are significantly more metal-rich than they were claimed in the literature.
\end{enumerate}

% --------------------------------------
\begin{acknowledgements}
% --------------------------------------

Authors are pleased to thank A.~Burenkov for the help in observations
with the 6\,m telescope. The suggestions of the referee F.Legrand helped
to improve the presentation of the data.
This work was partly supported by the INTAS grant 97-0033.
This research has made use of the
NASA/IPAC Extragalactic Database (NED) which is operated by the Jet
Propulsion Laboratory, California Institute of Technology, under contract
with the National Aeronautics and Space Administration.
The use of the Digitized Sky Survey from the Space Telescope Science
Institute was very helpful on various stages of this research.

\end{acknowledgements}

\end{document}